\documentclass{article}

\usepackage{arxiv}

\usepackage[utf8]{inputenc} % allow utf-8 input
\usepackage[T1]{fontenc}    % use 8-bit T1 fonts
\usepackage{hyperref}       % hyperlinks
\usepackage{url}            % simple URL typesetting
\usepackage{booktabs}       % professional-quality tables
\usepackage{amsfonts}       % blackboard math symbols
\usepackage{nicefrac}       % compact symbols for 1/2, etc.
\usepackage{microtype}      % microtypography
\usepackage{lipsum}

\usepackage{graphicx,subfig}%,widetext
\usepackage{cuted,multirow,comment,cite}
\usepackage{adjustbox}
\usepackage{mdframed,fancyvrb}
\usepackage{listings,amsmath}
\usepackage{multicol,ulem}

\title{On the validity of classical partition function.}

\author{
  Sushil K.~Singh \\
  Department of Physics \\SGTB Khalsa College\\
  University of Delhi
   \And
 Savinder ~Kaur\thanks{sk\_savinder2005@yahoo.co.in} \\
  Department of Physics\\ SGTB Khalsa College\\
  University of Delhi
}

\begin{document}
\maketitle

\begin{abstract}
The partition function for a system of non-interacting $N-$particles can be found by summing over all the states of the system. The classical partition function for an ideal gas differs from Bosonic or Fermionic partition function in the classical regime. Students find it difficult to follow the differences arising out of incorrect counting by the classical partition function by missing out on the indistinguishability of particles and Fermi-Bose statistics. We present a pedagogical computer-based experiment to probe and demonstrate the key differences in the partition functions (i) $Q$, for the system of distinguishable and indistinguishable particles (ii) $B$, for Bosons and (ii) $F$, for Fermions without formally using the single-particle partition function. 

\end{abstract}

% keywords can be removed
\keywords{Partition Function \and Gibbs Gas \and Fermions and Bosons}

%\begin{multicols}{2}
\section{Introduction}

The quintessential role that partition function (PF) plays in finding properties of a system is well established. The indistinguishability of identical particles has been a central point in the understanding of thermodynamic properties of the system. The thermodynamics function for a classical system can be obtained by defining the appropriate partition function
$$ Q^*_N \equiv \frac{1}{A} Q_N $$ 
where $A$ is a normalization factor and $Q_N = \int e^{\beta E_s}$ is the PF for  indistinguishable $N$-particles. The conclusion that entropy should be an additive function for macroscopic objects, led Gibbs to propose that the constant $A$ should be proportional to $N!$ in order not to overcount configurations of identical particles which are identical except for how the particles are labeled \cite{Berlinsky2019, Landau-Lif}. Though, the idea is totally unintelligible within classical mechanics, because interchanging particles does lead to classically different states, much has been deliberated on the appropriate normalization factor $A$ \cite{Buchdahl1974, Ford1971, Kroemer1980, Baierlein1997}. 
For most purposes, classical physics suffices to describe a system of $N$-particles under laboratory conditions and the indistinguishability is often incorporated by a judicious division by $N!$ \cite{McQuarrie1976, Kittel1980}. The factor $N!$, although correctly accounts for all states in which no orbital is occupied by more than one atom, fails for mutiple occupancy \cite{Kroemer1980}. We are tempted to summarize this discussion by saying that the number of quantum states for identical Fermions $n_F$ and that for identical Bosons $n_B$ are given by $n_F=n_B=n_C/N!$ where $n_C$ is the number of states for identical classical atoms. The division by $N!$ overcounts Fermi states and undercounts Bose states with the consequence that the PFs \cite{Kroemer1980, Baierlein1997}
$$ B_N > Q^*_N > F_N $$
where $B_N$ and $F_N$ represent the PF for $N$ non-interacting Bosons and Fermions respectively.
Though  the topic has been explored in detail \cite{Kroemer1980, Ford1971,  Baierlein1997}, the subtlety of the subject demands exploration and implementation in the computational lab. Despite the significance of this result, a pedagogical solution is not explicitly present in the literature. Our goal is to perform a {\it computer based experiment} (written in open source software Python) to analyse the differences between PFs, $Q_N$, $B_N$, $Q^*_N$, $F_N$ and the state functions which are usually overlooked in textbooks.
\section{Canonical Ensemble}

Consider a system of $N$-particles occupying a volume $V$ maintained at a constant temperature $T$, through thermal contact with a heat bath. A macrostate for the system can be realized by a large number of microstates with {\it equal a priori probabilities}. The microstate is defined as any possible individual configuration in the phase space. The system resides in any of these microstates at an instant $t$ while continually switching from one microstate to another. The more the number of available microstates the more the system spends time in the corresponding macrostate. The state functions are then expected to be "averaged" over these microstates through which the system passes \cite{Pathria2011}. An entirely different approach would be to imagine all these microstates to be occupied by {\it copies} of the system at the same instant. The average behavior of the system in this collection or {\it ensemble} would be identical to the time-averaged behavior of the given system. Statistically, the system evolves towards a macrostate with the largest number of microstates and spends an "overwhelmingly" large fraction of its time in this macrostate \cite{Berlinsky2019, Pathria2011, McQuarrie1976}.

\subsection{The “Canonical” Distribution Function}

For the system in equilibrium with a thermal reservoir at temperature $T$, the Boltzmann factor $exp(-\beta E_s)$ determines the “canonical” probability distribution $P_s$ for state $s$ 
\begin{equation}
P_s = \frac{e^{-\beta E_s}}{\sum_r^{all states} e^{-\beta E_r}}
\end{equation}
where $s,r$ refer to complete orthonormal set of energy eigenstates of the system. The properties of the reservoir appear only in the scalar factor $\beta=1/k_BT$ where $k_B$ is the Boltzmann's constant.

\subsection{The Partition Function}
The general recipe to extract information about the various macroscopic properties of the given system is to find the quantity $Q_N(V,T)$
%, the partition function of the system of particles confined to a volume $V$ at a temperature $T$ specified by $\beta$. It is also 
called the “sum-over-states” (German: Zustandssumme) or the partition function
\begin{equation}
Q_N(V,T) = \sum_s^{\textrm{all states}} e^{-\beta E_s}
\end{equation}
The dependence on $N$ and $V$ comes through the energy eigenvalues $E_s$. In most physical cases the energy levels {\it accessible} to a system are degenerate, that is, one has a group of states, $g_s$ in number, all belonging to the same energy value $E_s$. In such cases it is more useful to write the partition function as
\begin{equation}
Q_N(V,T) = \sum_s^{\textrm{energy states}} g_s e^{-\beta E_s}
\end{equation}
{\it The partition function counts the number of states available to the system.} 

\noindent The connection with thermodynamics is made by the Helmholtz Free energy $(H=E-TS)$
\begin{equation}
H_N(V,T) = -\frac{1}{\beta} \ln Q_N
\end{equation}
which is used to determine all other state functions of the system.

\section{Computer Based Experiment (CBE)}
Initially, we consider a system of {\it distinguishable} $N$-particles in which each particle can access $n$ number of single-particle energy levels $\epsilon_r$. There are $M=n^N$ possible combination of energy states available to the system. Each state of the system is characterized by energy $E_s$ ($s=1,...,M$)
\begin{equation} 
E_s = \epsilon_{r_1}^s + \epsilon_{r_2}^s + \dots + \epsilon_{r_N}^s
\end{equation}
where $\epsilon_{r_m}^s$ ($m=1,...,N$ and $r=1,...,n$) denotes energy of single-particle state when the $m^{th}$ particle is in its $r^{th}$ level while the system is in the $s^{th}$ state. The PF $Q_N$ can be written as
\begin{equation}
Q_N = \sum_{s=1}^{M} e^{-\beta\sum_{m=1}^{N} \epsilon_{r_m}^s} = \sum_{s=1}^{M} \left( \prod_{m=1}^{N} e^{-\beta \epsilon_{r_m}^s}\right)
\end{equation}
For {\it distinguishable} particles it is often useful to define a single-particle PF $q=\sum_r e^{-\beta \epsilon_r}$ where $\epsilon_r$  belongs to the complete orthonormal set of single-particle observable energies. The $N$-particle PF can then be found as $Q_N=q^N$. However, this single-particle PF $q$ is not a relevant quantity when we are dealing with quantum statistics, that is, Fermi-Dirac or Bose-Einstein statistics. The Fermionic and Bosonic particles are indistinguishable and are not independent because of the symmetry requirements of the wave functions \cite{Ford1971, McQuarrie1976}. We thus approach the problem without using the single-particle PF. We numerically calculate PFs for an  $N$-particle system with each particle being able to access $n$ energy levels. All the three cases (a) the distinguishable system (b) the Bosonic system and (c) the Fermionic system will be considered to find the differences arising in the number of available states. 

To illustrate the {\it CBE}, consider $2$ identical {\it distinguishable} particles each of which can access the $3$ energy levels labelled $0$ ($\epsilon_0$), $1$ ($\epsilon_1$) and $2$ ($\epsilon_2$). There are $3^2=9$ available states $S_p(i,j)$ to the system where $p$ refers to a state in  which the particle $1$ is in the $i^{th}$ energy level and particle $2$ is in the $j^{th}$ energy level. 
\begin{table}[ht]
\begin{center}
\begin{adjustbox}{width=0.45\textwidth}
\begin{tabular}{|llr|}
\multicolumn{3} {c}{(i) $(\epsilon_0, \epsilon_1, \epsilon_2)$}  \\
\hline
%{\tiny State}  & {\tiny Energy} & {\tiny Degeneracy} \\
$S_p(i,j)$  & $E_s$ & $g_s$ \\
\hline
$S_1 (0,0)$ & $E_0=2\epsilon_0$ & $1$ \\
\hline
$S_2 (0,1), S_3 (1,0)$ & $E_1=\epsilon_0+\epsilon_1$ & $2$ \\
\hline
$S_4 (0,2),S_5 (2,0)$ & $E_2=\epsilon_0+\epsilon_2$ & $2$ \\
\hline
$S_6 (1,1)$ & $E_3=2\epsilon_1$ & $1$ \\
\hline
$S_7 (1,2), S_8 (2,1)$ & $E_4=\epsilon_1+\epsilon_2$ & $2$ \\
\hline
$S_9 (2,2)$ & $E_5=2\epsilon_2$ & $1$  \\
\hline
\end{tabular}%
\begin{tabular}{|cc|}
\multicolumn{2} {c}{(ii) $(0, 1, 2)$}\\
\hline
%{\tiny Energy} & {\tiny Degeneracy} \\ 
$E_s$ & $g_s$ \\
\hline
\hline
$0$ & $1$ \\
\hline
$1$ & $2$ \\
\hline
\multirow{2}{*}{$2$} & \multirow{2}{*}{$3$} \\
 &  \\
\hline
$3$ & $2$ \\
\hline
$4$ & $1$ \\
\hline
\end{tabular}%
\end{adjustbox}
\end{center}
\caption{Distinguishable Particles : $S_p(i,j)$ observable states for 2-particle 3-level system, $E_s$ energy and $g_s$ degeneracy for levels (i) ($\epsilon_0$, $\epsilon_1$, $\epsilon_2$) \& (ii) in particular ($\epsilon_0=0$, $\epsilon_1=1$, $\epsilon_2=2$)}
\label{Tab:01}
\end{table}

The Table \ref{Tab:01} shows $9$ possible states of the system ($p=1,2...9$). States with label $2$ and $3$ i.e. $S_2(0,1)$ and $S_3(1,0)$ in which the particles just exchange their energy levels will be degenerate. Thus energy with label $1$ i.e. $E_1$ has degeneracy $g_1=2$. States with same energy are listed together and thus there are only $6$ distinct energy $E_s$ of the system (label $s$ is $0$ to $5$) with degeneracy $g_s$. 
\noindent Table \ref{Tab:01} also lists the allowed energy states when single-particle energy levels $\epsilon_0=0$, $\epsilon_1=1$ and $\epsilon_2=2$. Apart from the exchange degeneracy, the choice of $\epsilon$'s also leads to additional degeneracy in $E_2$ and $E_3$. 

\begin{table}[h]
%\resizebox{0.9\textwidth}{!}{%
\begin{minipage}{.5\textwidth}
\begin{center}
\begin{adjustbox}{width=0.85\textwidth}
\begin{tabular}{|llr|}
\multicolumn{3} {c}{(i) $(\epsilon_0, \epsilon_1, \epsilon_2)$}  \\
\hline
$S_p(i,j)$ & $E_s$ & $g_s$ \\
\hline
\hline
$S_1 (0,0)$ & $E_0=2\epsilon_0$ & $1$ \\
\hline
$S_2 (0,1) = S_3 (1,0)$ & $E_1=\epsilon_0+\epsilon_1$ & $1$ \\
\hline
$S_4 (0,2) = S_5 (2,0)$ & $E_2=\epsilon_0+\epsilon_2$ & $1$ \\
\hline
$S_6 (1,1)$ & $E_3=2\epsilon_1$ & $1$ \\
\hline
$S_7 (1,2) = S_8 (2,1)$ & $E_4=\epsilon_1+\epsilon_2$ & $1$ \\
\hline
$S_9 (2,2)$ & $E_5=2\epsilon_2$ & $1$  \\
\hline
\end{tabular}%
\begin{tabular}{|cc|}
\multicolumn{2} {c}{(ii) $(0, 1, 2)$}\\
\hline
$E_s$ & $g_s$ \\
\hline
\hline
$0$ & $1$ \\
\hline
$1$ & $1$ \\
\hline
\multirow{2}{*}{$2$} & \multirow{2}{*}{$2$} \\
 &  \\
\hline
$3$ & $1$ \\
\hline
$4$ & $1$ \\
\hline
\end{tabular}%
\end{adjustbox}
\end{center}
%}
\caption{ Same as Table \ref{Tab:01} but for Bosons }
\label{Tab:02}
\end{minipage}%
\begin{minipage}{.5\textwidth}
\begin{center}
\begin{adjustbox}{width=0.85\textwidth}
\begin{tabular}{|llr|}
\multicolumn{3} {c}{(i) $(\epsilon_0, \epsilon_1, \epsilon_2)$}  \\
\hline
FD $(\epsilon_1,\epsilon_2)$ & $E_s$ & $g_s$ \\
\hline
\hline
$S_1 (0,0)$ & $\times$ & $\times$ \\
\hline
$S_2 (0,1) = S_3 (1,0)$ & $E_1=\epsilon_0+\epsilon_1$ & $1$ \\
\hline
$S_4 (0,2) = S_5 (2,0)$ & $E_2=\epsilon_0+\epsilon_2$ & $1$ \\
\hline
$S_6 (1,1)$ & $\times$ & $\times$ \\
\hline
$S_7 (1,2) = S_8 (2,1)$ & $E_4=\epsilon_1+\epsilon_2$ & $1$ \\
\hline
$S_9 (2,2)$ & $\times$ & $\times$  \\
\hline
\end{tabular}%
\begin{tabular}{|cc|}
\multicolumn{2} {c}{(ii) $(0, 1, 2)$}\\
\hline
$E_s$ & $g_s$ \\
\hline
\hline
$\times$ & $\times$ \\
\hline
$1$ & $1$ \\
\hline
$2$ & $1$ \\
\hline
$\times$ & $\times$ \\
\hline
$3$ & $1$ \\
\hline
$\times$ & $\times$ \\
\hline
\end{tabular}%
\end{adjustbox}
\end{center}
%}
\caption{ Same as Table \ref{Tab:01} but for Fermions }
\label{Tab:03}
\end{minipage}%
\end{table}

The Table \ref{Tab:02} lists states available for the same system but with indistinguishable Bosonic particles along with energy eigen-states $E_s$ and their degeneracies $g_s$. The symbols have the same meaning as in Table \ref{Tab:01}. Since the Bosons are indistinguishable, states $S_2(0,1)$ and $S_3(1,0)$ are now the same state with just two different labels. This has to be counted only once as exchange of particle for indistinguishable particles is meaningless. Thus the degeneracy $g_s$ for corresponding energy $E_1$ is now counted as $g_1=1$ and not $2$. Thus all the exchange degeneracy are removed. However, the choice of $\epsilon_0=0$, $\epsilon_1=1$ and $\epsilon_2=2$ leads to a degeneracy in energy states $E_2$ and $E_3$.

\noindent The Table \ref{Tab:03} lists states available for the same system but with indistinguishable Fermions along with energy eigen-states $E_s$ and their degeneracies $g_s$. Fermions are indistinguishable particles and have additional restriction that no two particles occupy the same single-particle state (Pauli Exclusion principle). Thus the exchange degeneracy are again removed and the Pauli Exclusion principle disqualifies some states ($S_1$, $S_6$ \& $S_9$) as being observable. Degeneracy may arise due to choice of $\epsilon$ for a larger system.

The PF $Q$, $B$ and $F$ for the three systems are then evaluated. The PF for the Gibbs gas is also evaluated as $Q^* = Q/N!$. The program is provided in the Appendix A.  

\section{Results} 

\subsection{Partition Function}

\begin{figure}[ht]
\centering
\begin{minipage}{.5\textwidth}
\centering
\includegraphics[scale=.6]{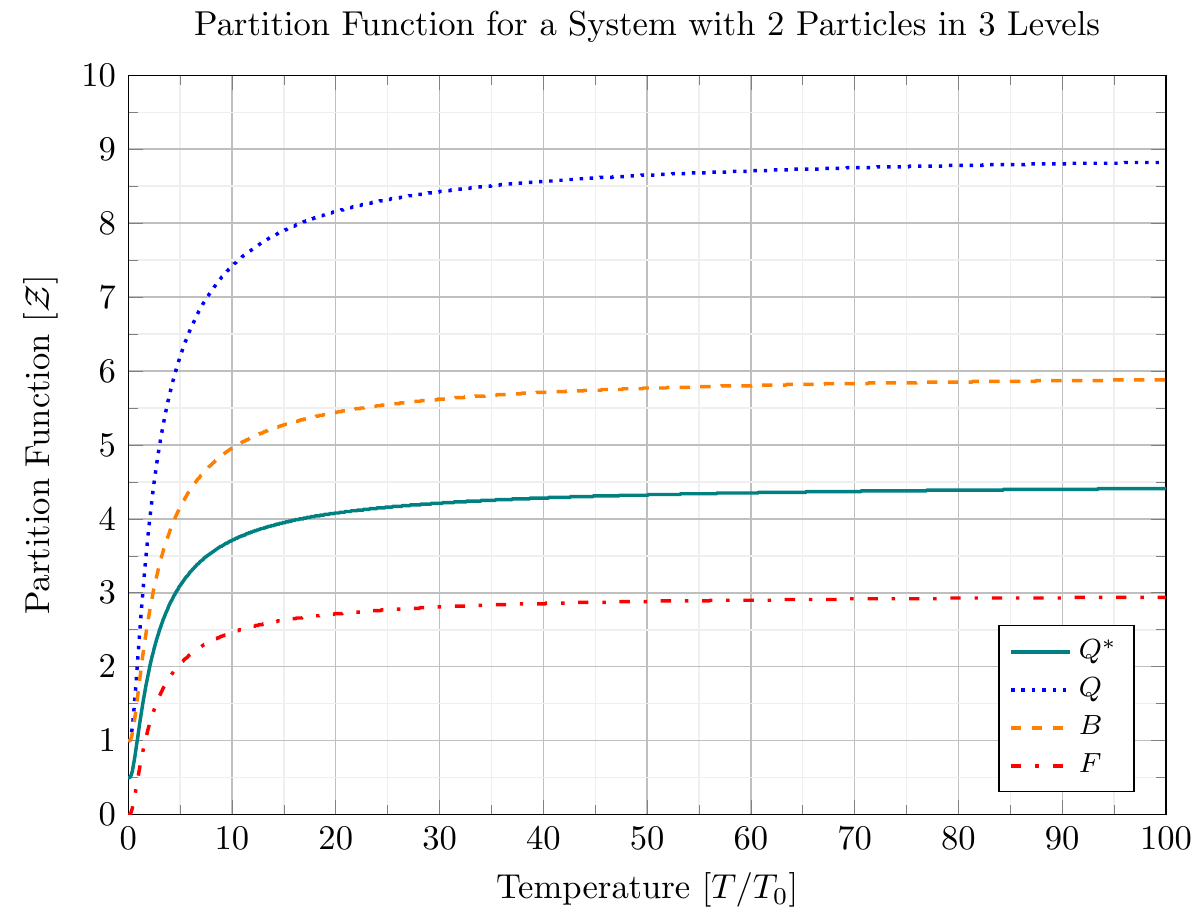} 
\caption{$2$-particle $3$-level Partition Function}
\label{fig: 01}
\end{minipage}%
\begin{minipage}{.5\textwidth}
\centering
\includegraphics[scale=.6]{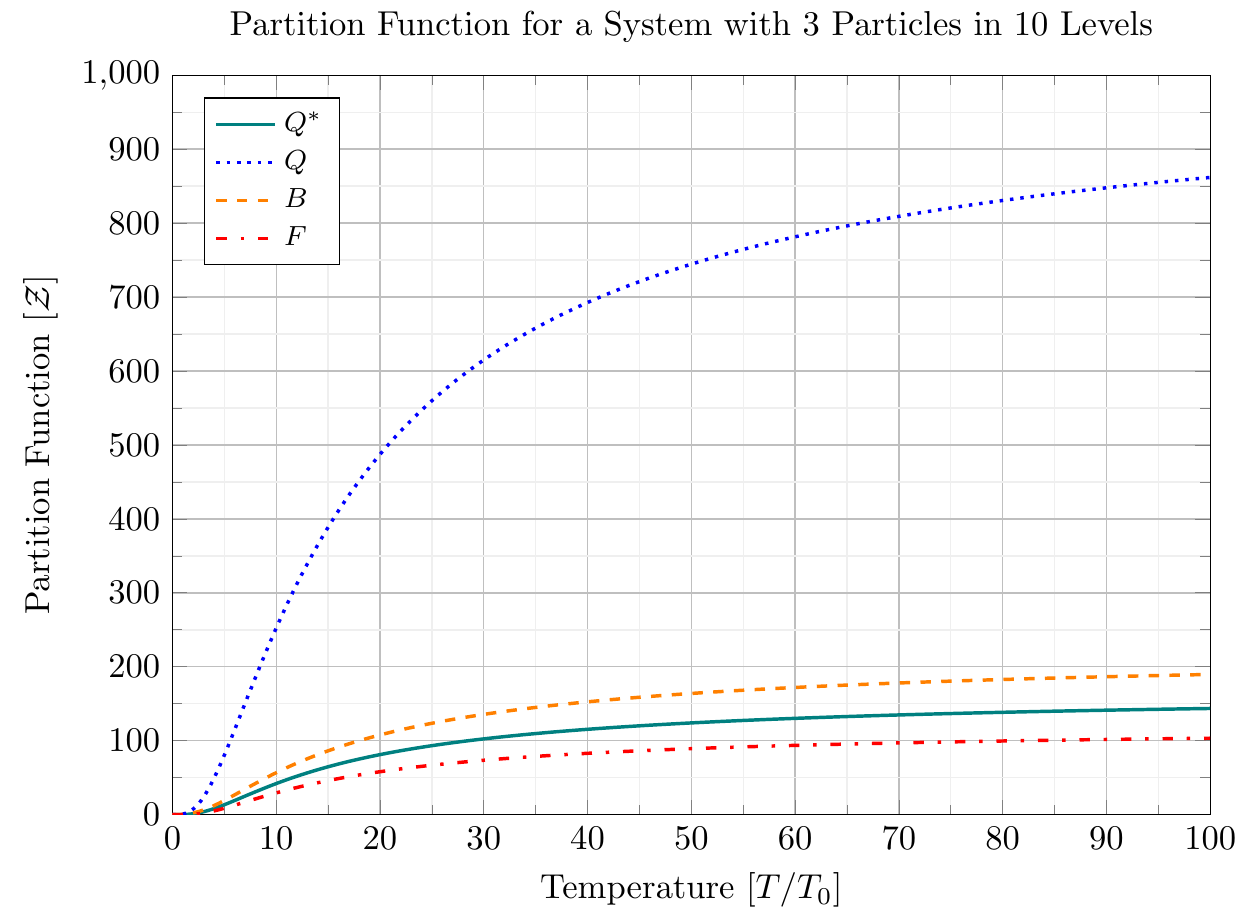} 
\caption{$3$-particle $10$-level Partition Function}
\label{fig: 02}
\end{minipage}%
\end{figure}

The PFs for $2$-particle $3$-level distinguishable system $Q_N$, Gibbs gas $Q^*_N$, Boson $B_N$ and Fermion $F_N$ are plotted with respect to dimensionless temperature $T/T_0$ where $T_0=\epsilon_0/k_B$ in Figure \ref{fig: 01}.
At high $T/T_0$ values (i) $Q\rightarrow 9$ for distinguishable particles, (ii) $B\rightarrow 6$ for Bosonic particles, (iii) $F\rightarrow 3$ for Fermionic system and (iv) $Q^*\rightarrow 4.5$ for Gibbs gas. They reflect the available energy states as can be verified from the respective Tables \ref{Tab:01}, \ref{Tab:02} and \ref{Tab:03}. Further, for all temperatures the partition function for Gibbs gas follows the inequality $B_N>Q^*_N>F_N$ \cite{Kroemer1980, Baierlein1997}. 

In figures $2-6$ we consider a modest $3$-particles and $10$-level system which has $10^3$ accessible energy states. Computationally, we need finite particles and levels to study the behaviour as we scale up the system. The $1000$ available states are adequate to show that the $1/N!$ term is justified as we choose larger number of particles. 

The PF for $3$-particle $10$-level distinguishable system, Gibbs gas, Boson and Fermion are shown in Figure \ref{fig: 02}. Though the system is still small, the factor of $1/N!$ used to scale down the distinguishable system PF $Q$ to that for a Gibbs gas $Q^*$ seems to predict values in-between those of $B$ and $F$.

\subsection{State Functions}
\begin{figure}[ht]
\centering
\begin{minipage}{.5\textwidth}
\centering
    \includegraphics[scale=.6]{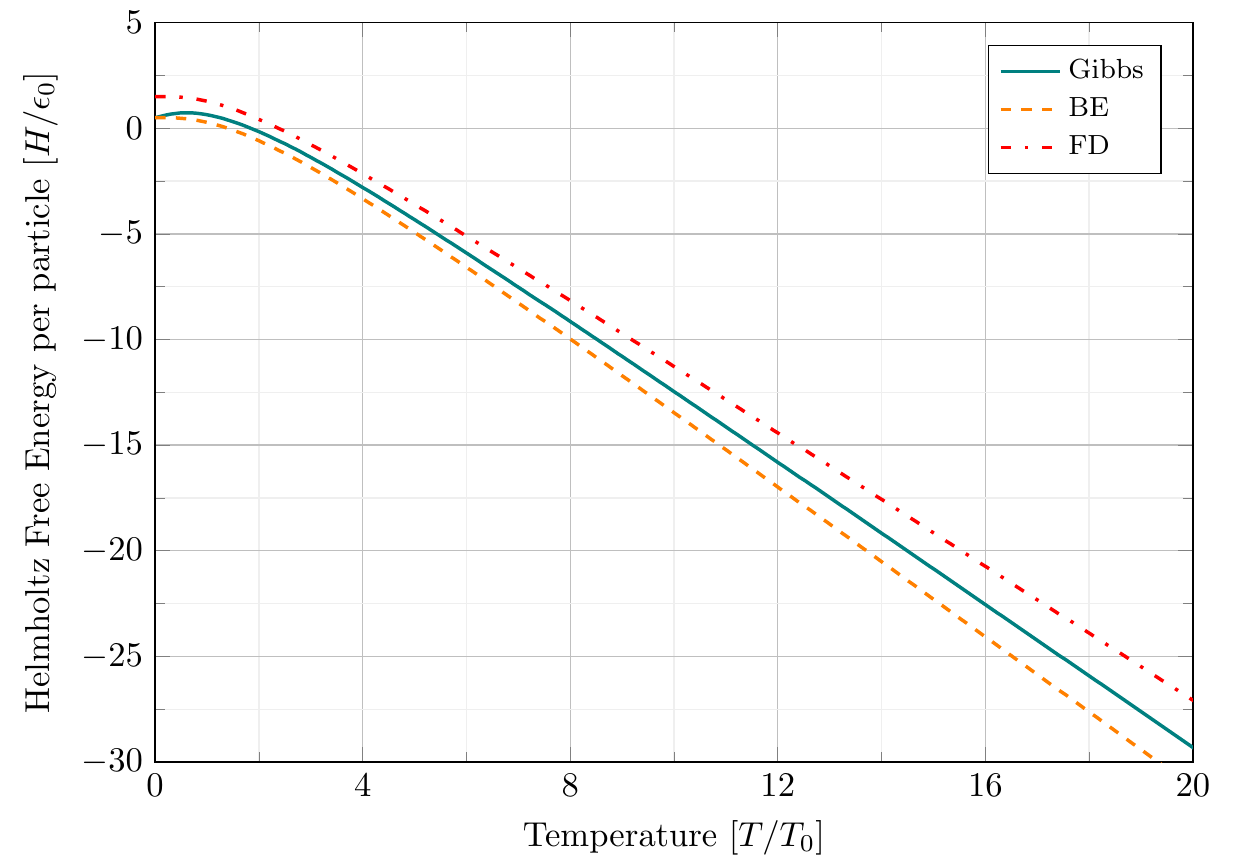} 
    \caption{The Helmholtz Free Energy}
\label{fig: 03}
\end{minipage}%
\begin{minipage}{.5\textwidth}
\centering
    \includegraphics[scale=.6]{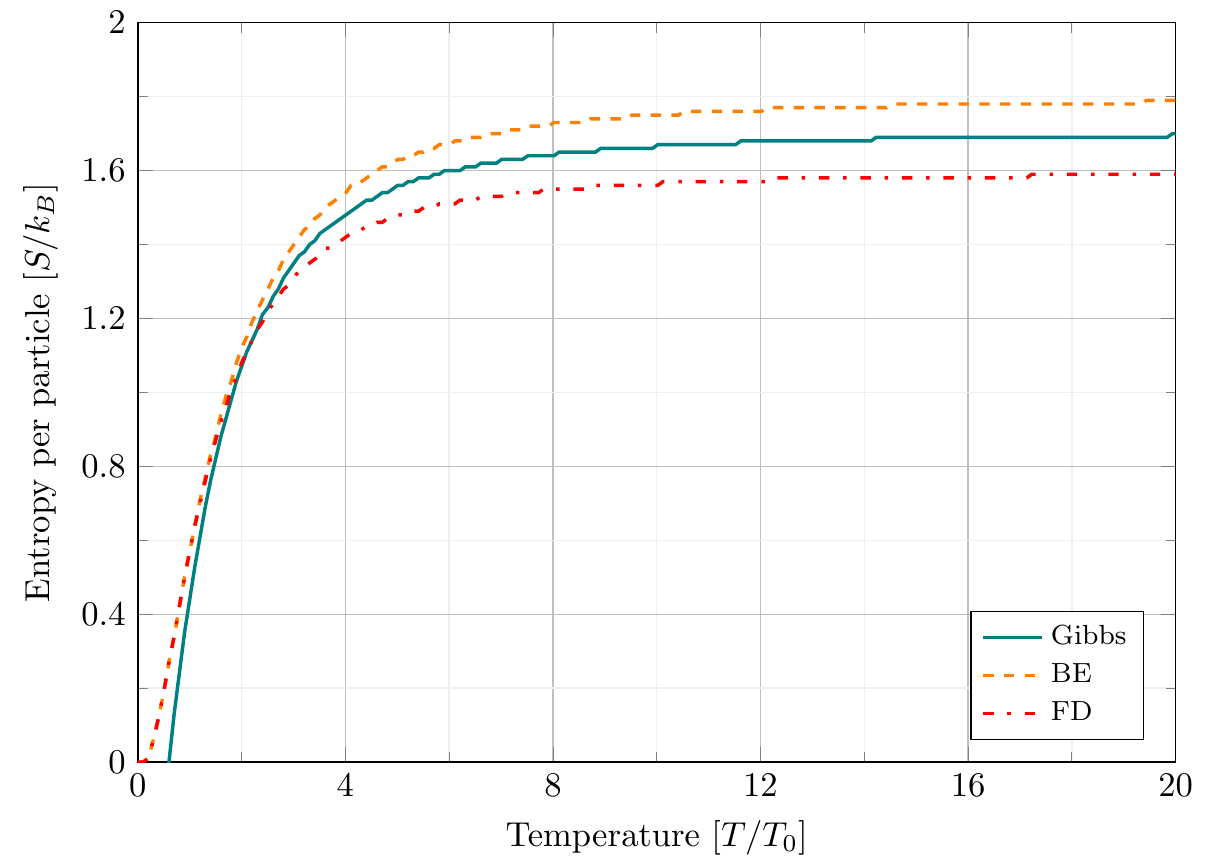} 
    \caption{The Entropy}
\label{fig: 04}
\end{minipage}%
\end{figure}

\begin{figure}[ht]
\centering
\begin{minipage}{.5\textwidth}
\centering
    \includegraphics[scale=.6]{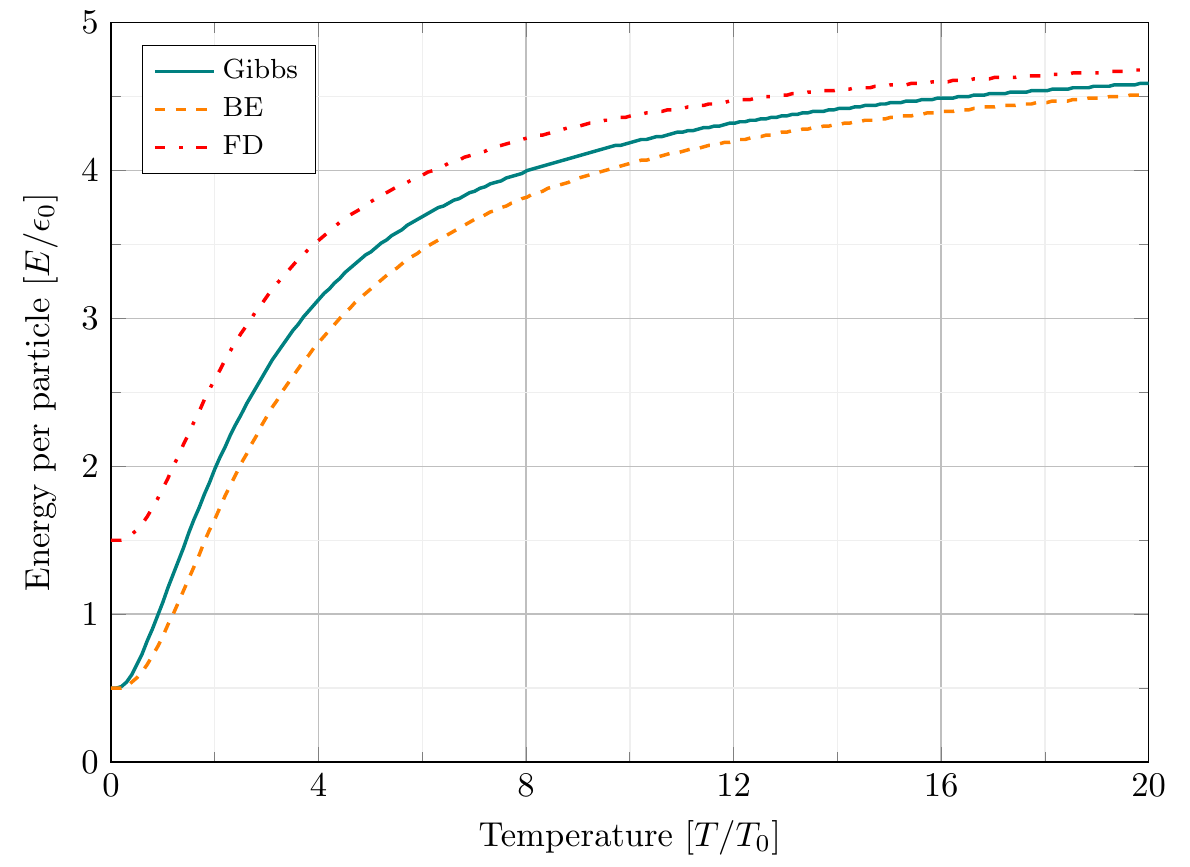} 
    \caption{The Energy}
    \label{fig: 05}
\end{minipage}%
\begin{minipage}{.5\textwidth}
\centering
    \includegraphics[scale=.6]{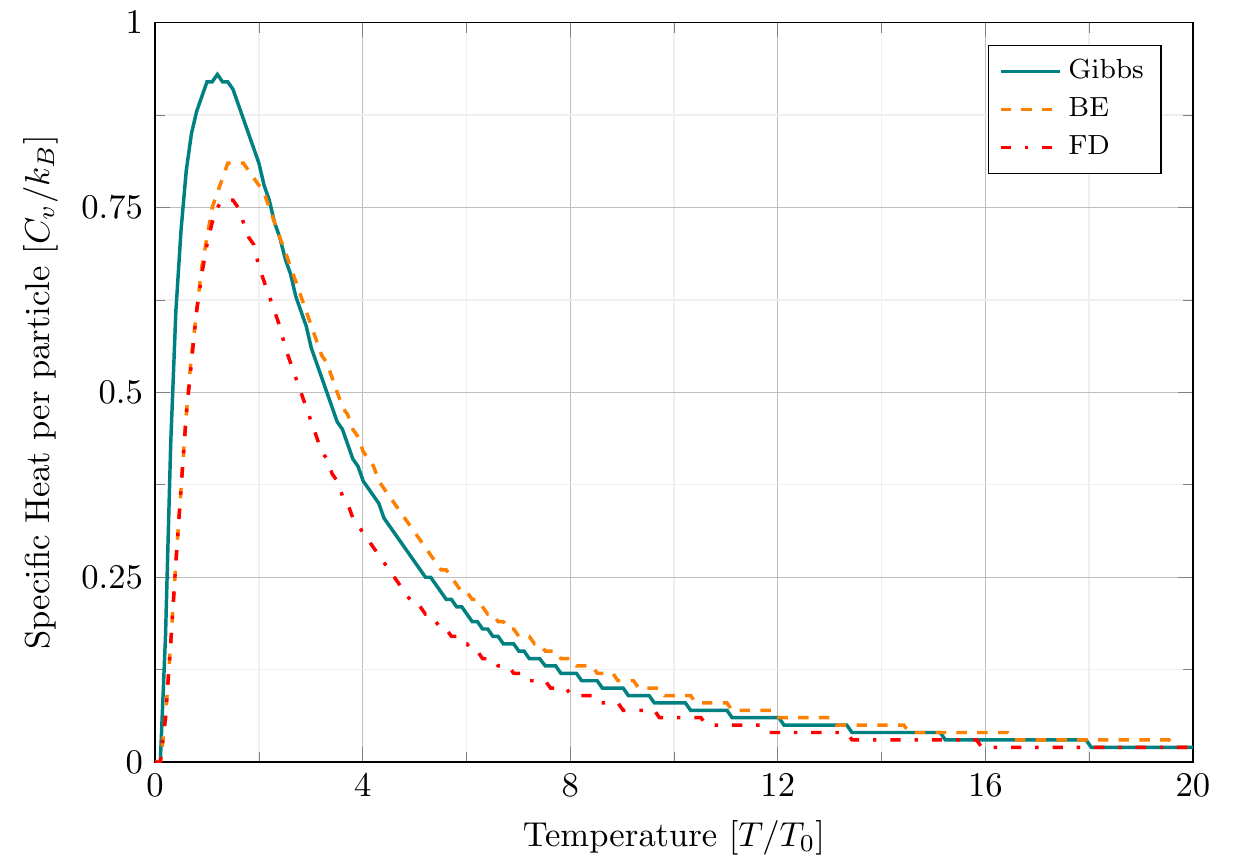} 
    \caption{The Specific heat} \label{fig: 06}
\end{minipage}%
\end{figure}

For the $3$-particle $10$-level system (with $\epsilon_0=1$, $\epsilon_1=2$, $\dots$ and $\epsilon_9=10$) the non-dimensional state functions $H/\epsilon_0$, $S$, $E/\epsilon_0$ and $C_v/\epsilon_0$ per particle are presented with $T/T_0$ in Figures \ref{fig: 03}, \ref{fig: 04}, \ref{fig: 05} and \ref{fig: 06} respectively. 
Though the system is still small, the factor of $1/N!$ used to scale down the distinguishable system partition function $Q$ to that for a Gibbs gas $Q^*$ seems to predict values of the state functions close to those for $B$ and $F$. We see that the logarithmic dependence of the state function on the partition function bring the Gibbs gas closer to Boson and Femion gas \cite{Landau-Lif}.

\section{Conclusions}
We find that (a) the PFs $Q_N^*$, $B_N$ and $F_N$ are different and (b) by taking large number of particles with even larger number of energy states, the state functions for Bosons and Fermions approach those for the classical case because the state functions have logarithmic dependence on the PF.

The advantage of this approach is that 
(i) students by ignoring the single-particle PF understand states available to the system itself 
(ii) the PF correctly counts the available states and provides an interpretation to the PF.
(iii) Even though the PFs $Q_N$, $B_N$ and $F_N$ differ the state functions can be described by using the $Q^*_N=1/N!\times Q_N$

%\section{Acknowledgements}

%We acknowledge SGTB Khalsa College, University of Delhi for providing the infrastructure.

%\section{References}
%%%%%%%%%%%%%%%%%%%%%%%%%%%%%%%%%%%%%%%%%%%%%%%%%%%%%%%%%%%%%%%%%%%%%%%%%%%%%%%

%\end{multicols}

\end{document}